\documentclass[prl,twocolumn]{revtex4-2}
\usepackage[bottom]{footmisc} 

\usepackage{graphicx}
\usepackage{subcaption}
\usepackage{amsmath}
\usepackage{mathrsfs}
\usepackage{xcolor}
\usepackage{hyperref}
\usepackage{cleveref}
\usepackage{comment}
\usepackage{gensymb}

\usepackage{dcolumn}
\usepackage{bm}

\hypersetup{colorlinks=true, linkcolor=blue, citecolor=blue, urlcolor=blue, filecolor=blue}

\begin{document}

\title{Axion Production and Detection Using a Dual NMR-type Experiment}

\author{Jeff A. Dror$^1$}
\author{Qiushi Wei$^1$}
\author{Fengwei Yang$^{1,2}$}

\affiliation{$^1$Institute for Fundamental Theory, Physics Department,
University of Florida, Gainesville, FL 32611, USA}

\affiliation{$^2$Department of Physics and Astronomy,
University of Notre Dame, South Bend, IN 46556, USA}

\begin{abstract}
Axions that couple to nuclear spins via the axial current interaction can be both {\em produced and detected} using nuclear magnetic resonance (NMR) techniques. In this scheme, nuclei driven by a real oscillating magnetic field in one device act as an axion source, which can drive NMR in a nearby spin-polarized sample interrogated with a sensitive magnetometer. We study the prospects for detecting axions through this method and identify two key characteristics that result in compelling detection sensitivity. First, the gradient of the generated axion field can be substantial, set by the inverse distance from the source. Near the source, it reduces to the inverse of the source’s geometric size. Second, because the generated axion field is produced at a known frequency, the detection medium can be tuned precisely to this frequency, enabling long interrogation times. We show that the experimental sensitivity of a pair of centimeter-scale NMR devices operating over a 15-day integration time can already surpass existing astrophysical bounds on the axion-nucleon coupling. A similar sensitivity can be achieved with 10 centimeter-scale NMR devices with only 1 hour of integration time. These dual NMR configurations are capable of probing a wide range of axion masses, up to values comparable to the inverse distance between the source and the sensor.

\end{abstract}

\maketitle

\noindent{\bf Introduction.}
{\it Axions} are derivatively coupled light scalar fields that appear in many extensions of the Standard Model. They provide a compelling solution to the strong CP problem, serve as viable dark matter candidates, and arise naturally in string theory~\cite{Peccei:1977hh,Weinberg:1977ma,Wilczek:1977pj,Preskill:1982cy,Abbott:1982af,Dine:1982ah,Svrcek:2006yi}. Their derivative couplings ensure that the axion mass is protected from large quantum corrections by a shift symmetry, broken only by non-perturbative effects. Consequently, axions remain light, and their interactions with the Standard Model are typically extremely weak.

It is conceivable that axions persist but do not make up a substantial cosmological abundance. If so, their detection must rely on the production of axions in the laboratory. In the case of axion-photon couplings, this has motivated experiments employing light-shining-through-walls experiments~\cite{VanBibber:1987rq,Hoogeveen:1992nq,Sikivie:2007qm,ALPS:2009des,Caspers:2009cj,Ehret:2010mh,Bahre:2013ywa,Betz:2013dza,Janish:2019dpr}, which utilize radio or optical frequency cavities with background electromagnetic fields. 
If an axion carries both derivative (CP-conserving) and non-derivative (CP-violating) interactions with nuclei, then it has been proposed to generate axions with mechanical oscillators or rotators and detect its presence using nuclear magnetic resonance (NMR) techniques~\cite{Arvanitaki:2014dfa,ARIADNE:2017tdd,Almasi:2018cob,Arvanitaki:2024dev}. For purely derivative couplings, existing searches for exotic spin-dependent interactions  using atomic magnetometers \cite{PhysRevLett.133.191801,Xu:2025lly} focus on the static limit where the axion mass is much larger than the characteristic oscillation frequency $(m_a\gg\omega)$.

In this {\it Letter}, we present a novel production and detection scheme for axions that relies solely on the axial current interaction between axions and nucleons using a pair of NMR-type experiments, providing the first comprehensive dynamical framework for laboratory-scale axion production. Independent of the ambient axion abundance, axion production is realized by an ``inverse'' version of an NMR-based axion detector, where the precessing spins of nuclei driven by an external alternating magnetic field generate an oscillating axion field, which can in turn be detected by an additional nearby NMR device.
This experimental setup extends current spin-precession axion experiments, such as CASPEr~\cite{Budker:2013hfa,Graham:2013gfa,Abel:2017rtm,JacksonKimball:2017elr,Wu:2019exd,Garcon:2019inh,Aybas:2021nvn}, to probe axions not sourced cosmologically.

Unlike axion dark matter, which oscillates at its mass and has a finite coherence time, the NMR-generated axion field is produced coherently at the nuclear spin precession frequency, independent of the axion mass. This allows a dual-NMR setup to probe a wide range of axion masses with a single resonance frequency and enables much longer interrogation times than typical dark matter searches. Moreover, the axion field gradient—which drives the NMR detector—is determined by the source geometry and distance, and can be significantly larger than the corresponding gradient in the dark matter case. Together, these features provide promising sensitivity to the axion-nucleon coupling $g_{aN}$, even though the generated field amplitude remains much smaller than that of ambient axion dark matter.

In this {\it Letter}, we will derive both the production and response in NMR devices, present the projected sensitivities for specific experimental setups, and examine their dependence on experimental parameters.

\begin{figure*}[h!t]
        \centering
         \includegraphics[width=0.47\linewidth]{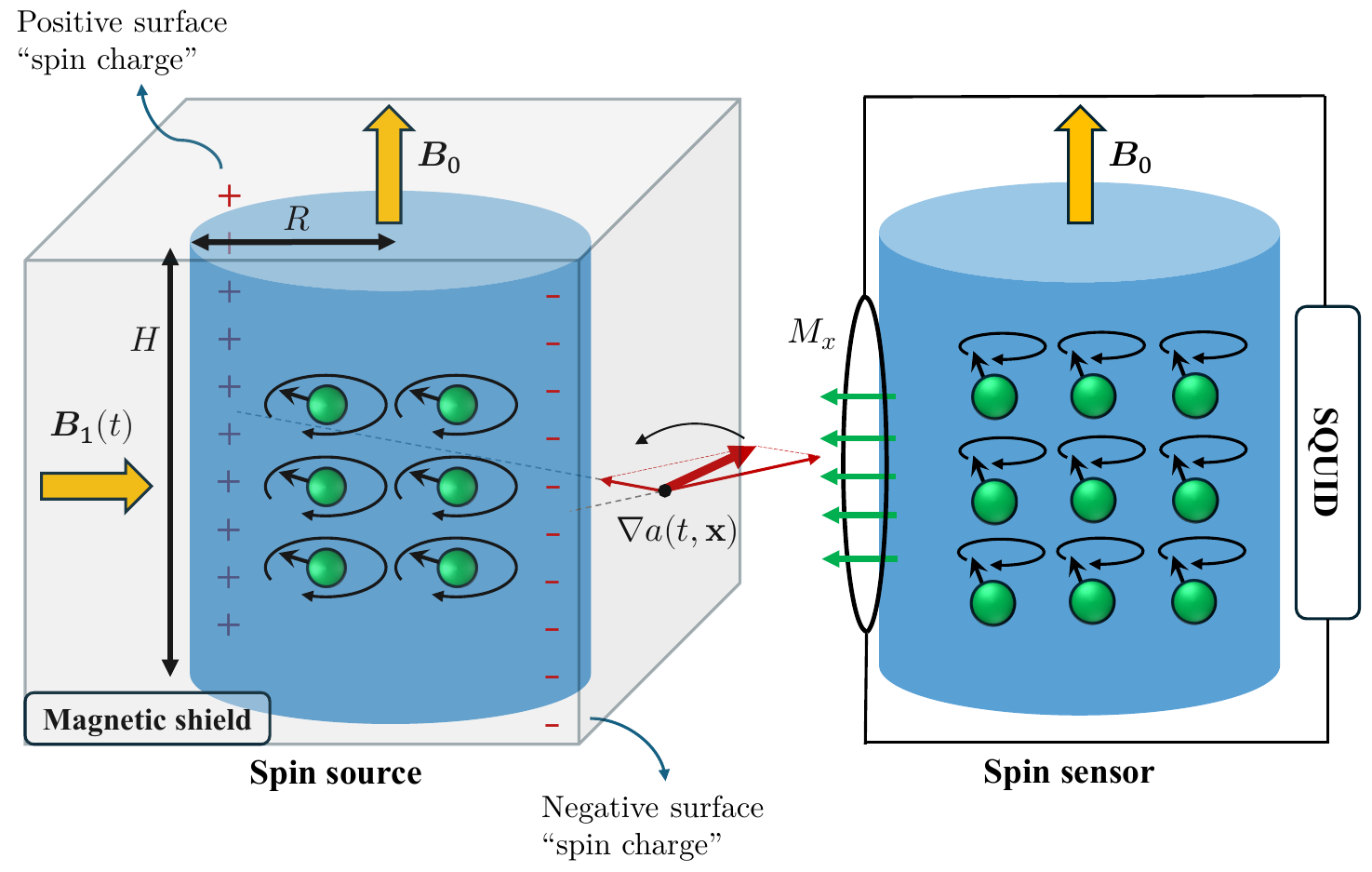}
         \qquad\includegraphics[width=0.48\linewidth]{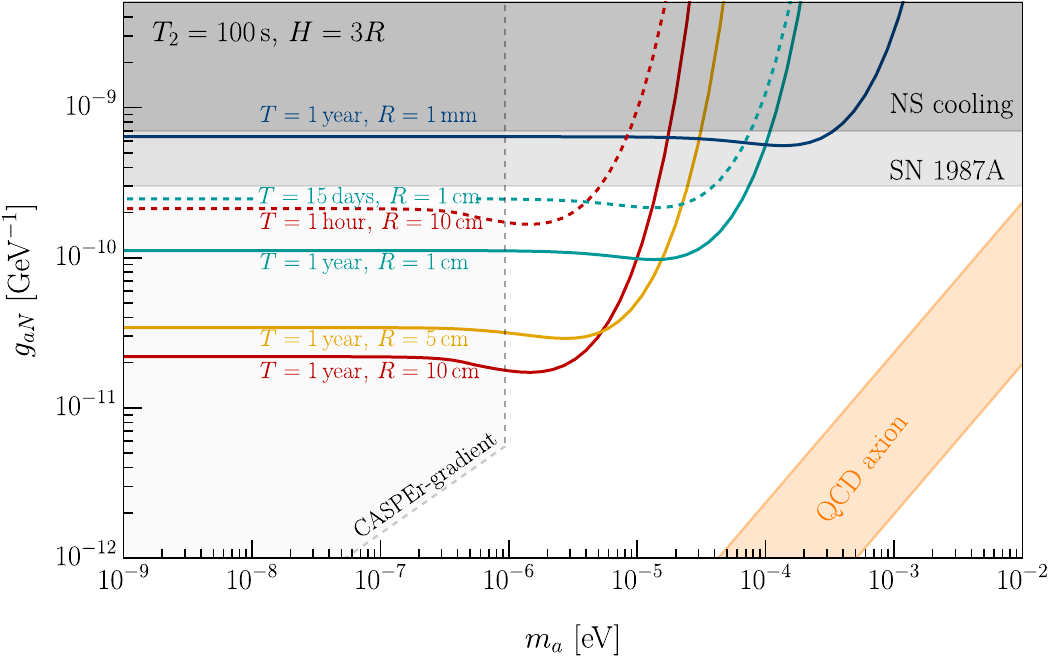}
    \caption{Summary of the experimental proposal. {\bf Left:} The sketch of the experiment. {\bf Right:} The projection of the 95\% C.L. upper limit of axion-nucleon coupling $g_{aN}$ as a function of axion mass $m_a$. The material in both the spin source and spin sensor is $^{129}\mathrm{Xe}$, with a number density $n_N=1.3\times10^{22}\,{\rm cm}^{-3}$ and a Larmor frequency $\omega_0 =0.74\mathrm{GHz}$ which corresponds to a static magnetic field of $B_0=10\,\mathrm{T}$, and we take the transverse relaxation time to be $T_2=100$\,s. For $^{129}$Xe, $g_{aN}\simeq 0.600g_{an}+0.046g_{ap}$, where $g_{an}$ and $g_{ap}$ are axion-neutron and axion-proton couplings (see Supplemental Material for details). The source and the detector cylinders radii are varied by $R=10\,$cm, 5\,cm, 1\,cm, and 1\,mm, with their heights fixed at $H=3R$. The two cylinder axes are separated by $3R$ and the pickup loop is placed at the near end of the sensor medium. The observation time for the solid lines is $T=1\,\mathrm{yr}$, while the dashed lines use the same setups as the solid lines in the same colors, but run for different observation times. The astrophysical bounds on $g_{aN}$ from neutron star (NS) cooling~\cite{Buschmann:2021juv} and SN 1987A~\cite{Lella:2023bfb} are shown in gray-shaded regions, same as those taken in Ref.~\cite{Dror:2022xpi}. The QCD axion parameter space is shown in the orange-shaded region. For comparison, the CASPEr-gradient projection~\cite{JacksonKimball:2017elr} is overlaid, which relies on the assumption of axions as 100\% dark matter. }
    \label{fig:sensitivity}
\end{figure*}

\vspace{0.2cm}
\noindent{\bf Axion Field Production.\label{sec:production}}
We start with the production of the axion field from a spin-precessing source. 
The axial current interaction between an axion $a$ and a nucleon $N$ is~\footnote{Axions can also couple to the electric dipole moment of nucleons through the operator $-ig_da\bar{N}\sigma_{\mu\nu}\gamma^5NF^{\mu\nu}/2$. However, this coupling is stringently constrained by astrophysical observations, and we estimate that the sensitivities of the configurations discussed in this work are insufficient to surpass these constraints.}:
\begin{equation}\label{lagrangian}
    \mathcal{L}\supset  g_{aN}\partial_\mu a \bar{N}\gamma^\mu\gamma^5N.
\end{equation}
This interaction gives the field equation for axions:
\begin{equation}\label{EOM}
    (\partial_t^2 - \nabla^2+m_a^2)a(t,{\bf x})= -g_{aN} \partial_\mu j^{5\mu} (t,\mathbf{x}),
\end{equation}
where $j^{5\mu}=\bar{N}\gamma^\mu\gamma^5N$ is the nucleon axial current.
For a non-relativistic source medium with spin-1/2 nuclei, $j^{5i}$ is twice the nuclear spin density, while the temporal component $j^{50}$ is negligible.
The source term is thus a ``spin charge'' distribution induced by the spin polarization of the medium, $\rho_\mathrm{s}(t,\mathbf{x}) \equiv -\partial_i j^{5i}(t,\mathbf{x})$, exciting the axion field with a coupling strength $g_{aN}$ (see Supplemental Material for more discussions).

Supposing the source medium is uniform and the nuclear spins are precessing in alignment, then $j^{5i} (t,\mathbf{x}) \simeq n_N {\sigma}^{i}(t)$ inside the source, with $n_N$ the nuclei density and $\boldsymbol{\sigma}(t)$ a unit vector pointing towards the direction of nuclear spins at time $t$. 
In this case, the ``spin charge'' only distributes at the boundary of the source medium as surface charge, $\Sigma_\mathrm{s}(t,\mathbf{x}) \equiv n_N\,\hat{\bf n}\cdot\boldsymbol{\sigma}(t)$, where $\hat{\bf n}$ is the unit normal vector to the surface of the source at position $\mathbf{x}$.

If the nuclear spins precess coherently at frequency $\omega_0$, the oscillating ``spin charge'' distribution will generate a monochromatic oscillating axion field at the same frequency. We separate temporal and spatial dependence using a complex representation, writing $\boldsymbol{\sigma}(t) = e^{-i\omega_0t} \boldsymbol{\sigma}$ and $a(t,\mathbf{x})=e^{-i\omega_0t}a(\mathbf{x})$ where the physical quantities are obtained by taking the real parts. 

The axion field produced by such a source, solved from Eq.~\eqref{EOM}, is:
\begin{equation}
\label{eq:ax}
    a({\bf x}) = g_{aN}n_N \oint d^2 x' \frac{\hat{\bf n}'\cdot \boldsymbol{\sigma}}{4\pi|{\bf x}-{\bf x'}|}  e^{ik_0|{\bf x}-{\bf x'}|},
\end{equation}
where $k_0 \equiv \sqrt{\omega_0^2 - m_a^2}$ (if $\omega_0 \geq m_a$) or $i\kappa_0$ with $\kappa_0 \equiv \sqrt{m_a^2 - \omega_0^2}$ (if $\omega_0 < m_a$), $\hat{\bf n}'$ is the unit normal vector to the source surface at $\mathbf{x}'$, and the integration is performed over the source surface. In writing down Eq.~\eqref{eq:ax}, we assumed $\omega_0>0$ as it no longer represents the retarded solution if $\omega_0<0$ (see Supplemental Material for details).
The axion solution is divided into two regimes, depending on the relative magnitudes of the frequency $\omega_0$ and axion mass $m_a$.
For small axion masses ($m_a\le \omega_0$), the axion field is a propagating wave outward from the ``spin charge'' with a wavenumber $k_0$ and an amplitude decreasing with distance.
For large axion masses ($m_a>\omega_0$), the exponential decay factor $e^{-\kappa_0|{\bf x}-{\bf x'}|}$ indicates that the field is evanescent, decaying exponentially over a distance $1/\kappa_0$.

If the size of the source as well as the source-detector distance are both small enough such that $\omega_0|{\bf x}-{\bf x'}|\ll 1$, for all masses $m_a\lesssim1/R$ the axion field satisfies the near-field limit ($k_0|{\bf x}-{\bf x'}|\ll 1$ or $\kappa_0|{\bf x}-{\bf x'}|\ll 1$) and reduces to
\begin{equation}
\label{eq:a_near}
    a({\bf x})= g_{aN}n_N\oint d^2 x'\frac{\hat{\bf n}'\cdot \boldsymbol{\sigma} }{4\pi|{\bf x}-{\bf x'}|}.
\end{equation}
In this regime, $a(\mathbf{x})$ is independent of its mass $m_a$ and the precession frequency $\omega_0$ (except for the harmonic time dependence factor we factor out). 
The produced axion field can be considered as sourced instantaneously by the ``spin charge'' distribution.
For a GHz precession frequency, this can be achieved with centimeter-scale NMR experiments.

The near-zone field, after including the harmonic time dependence back into Eq.~\eqref{eq:a_near}, can be intuitively understood by drawing an analogy with electrostatics.
For any instant $t$, the axion field and its gradient are analogous respectively to the electric potential $\Phi$ and the negative electric field $-\mathbf{E}$ induced by an static electric charge distribution identical to the ``spin charge'' distribution, $\Sigma_\mathrm{s}(t,\mathbf{x}) = n_N\,\hat{\bf n}\cdot\boldsymbol{\sigma}(t)$, at that moment.
This correspondence is particularly useful for understanding the configuration of $\nabla a(t,\mathbf{x})$ in the near zone which drives the NMR in the detection medium --- at any moment, positive ``spin charge'' draws a component that points towards it, while negative ``spin charge'' leads to a component that points away from it, as sketched in Fig.~\ref{fig:sensitivity} ({\bf Left}).

As the axion source, we consider a cylindrical system as shown in Fig.~\ref{fig:sensitivity} ({\bf Left}), where the nuclear spins are polarized by a static magnetic field $\mathbf{B}_0=B_0\hat{\bf z}$ in the $+z$ direction and the NMR is driven by an alternating magnetic field $\mathbf{B}_1(t)$ in the $xy$ plane. The frequency of spin precession and the produced axion field $\omega_0$ is given by the Larmor frequency of the nuclei, $\omega_0 = |\gamma B_0|$, where $\gamma$ is the gyromagnetic ratio of the nuclei. We further assume the source sample is hyperpolarized and the nuclear spins maintain a 90\degree tilt and precess completely in the $xy$ plane, such that $\boldsymbol{\sigma} = \hat{\bf x} \mp i\hat{\bf y}$ ($-$ for $\gamma >0$ and $+$ for $\gamma <0$)~\footnote{This gives $\boldsymbol{\sigma}(t) = e^{-i\omega_0t}(\hat{\bf x}\mp i\hat{\bf y})$, which corresponds to the spins rotating in the clockwise (or counter-clockwise) direction around $\hat{\bf z}$ when viewed from above if $-$ (or $+$) is taken on, consistent with the dynamics of spins described by the Bloch equation for $\gamma >0$ (or $<0$). Note that in this complex representation, $\mathrm{Re}[ \boldsymbol{\sigma}(t)]$ is a unit vector.}.
The resulting near-field $\nabla a(t,\mathbf{x})$ rotates in the opposite direction as the bulk nuclear spins at the frequency $\omega_0$, inferred from the rotating surface ``spin charge'' on the lateral surface of the source cylinder.

For illustration of the magnitude of the produced axion field, we take the example of such a spin source with a height-to-radius ratio $H/R=3$ and its center located at the origin. We assume a number density of $n_{N} = 1.3\times10^{22}\,\mathrm{cm}^{-3}$, with both the field and the gradient scaling linearly with the density. Carrying out the numerical integral of Eq.~\eqref{eq:a_near}, the near-field solution outside the source and its gradient display a power-law behavior along the radial distance on the $xy$ plane:
\begin{align}
    &a(x\hat{{\bf x}}) \simeq 2.4\,\mu \mathrm{eV} ~\bigg(\frac{x}{R}\bigg)^{-1.6} \bigg(\frac{g_{aN}}{10^{-10}~\mathrm{GeV}^{-1}}\bigg)
     \bigg(\frac{R}{10~\mathrm{cm}}\bigg), \\
    &\partial_xa(x\hat{{\bf x}}) \simeq -7.6\, \mu \mathrm{eV}^2 \bigg(\frac{x}{R}\bigg)^{-2.6} \bigg(\frac{g_{aN}}{10^{-10}\mathrm{GeV}^{-1}}\bigg). 
\end{align}
Although the axion field is significantly smaller than that of any cosmological population of axion dark matter field, its gradient is relatively enhanced by $|\partial_x a/a| \sim 1/R \sim 1~\mu\mathrm{eV}(R/10~\mathrm{cm})^{-1}$ from its spatial inhomogeneity, which is about 4 orders of magnitude larger than that for axion dark matter with $\mu\mathrm{eV}$ masses~\footnote{For axion dark matter, the field amplitude is $a = \sqrt{2\rho_{\mathrm{DM}}}/m_a \sim 10^3\mathrm{eV}(m_a/1\mu \mathrm{eV})^{-1}$ and the field gradient is $|\nabla a| \sim m_a v_0a_0 \sim 10^{-6}\mathrm{eV}^2$. Additionally, $|\nabla a|/a \sim m_a v_0 \sim 10^{-9}\mathrm{eV}(m_a/1\mu \mathrm{eV})$. In these numerical expressions, we assumed an expected local density $\simeq0.3 \mathrm{GeV/cm^3}$ and a characteristic velocity of $v_0\sim10^{-3}$.}. 
Note that while the experimental setups discussed in this paper are around the near-field limit, we use the the full-form solution, Eq.~\eqref{eq:ax}, to make our final projections.

\vspace{0.2cm}
\noindent{\bf Axion Field Detection.\label{sec:detection}}
For a non-relativistic nucleus, the interaction Hamiltonian arising from the Lagrangian in Eq.~\eqref{lagrangian} is
\begin{equation}
    H_{\mathrm{int}} = -2 g_{aN} \, \nabla a(t,\mathbf{x}) \cdot \mathbf{S},
\end{equation}
where $\mathbf{S}$ is the nuclear spin operator. Comparison with the Zeeman interaction of spins in a magnetic field suggests interpreting the influence of the axion as inducing an effective magnetic field:
\begin{equation}\label{B_a}
    \mathbf{B}_a(t,\mathbf{x}) =  \frac{2 g_{aN}}{\gamma} \, \nabla a(t,\mathbf{x}).
\end{equation}
The oscillatory gradient of the axion field derived in the above section will generate this effective magnetic field that can drive another NMR device nearby as a spin sensor.

We consider a spin sensor analogous to the production source we discussed above -- a highly-polarized homogeneous medium with gyromagnetic ratio $\gamma$, engulfed in a uniform magnetic field $\mathbf{B}_0=B_0 \hat{\bf z}$ in the $+z$ direction. 
We assume the source and the detection medium have the same Larmor frequency $\omega_0$ so that the nuclear spins in the detection medium are resonantly driven.

Following a similar procedure to that in Ref.~\cite{Dror:2022xpi}, we solve the Bloch equation of the detection material driven by the axion-induced magnetic field $\mathbf{B}_a$ (see Supplemental Material for details).
The axion-induced transverse magnetization $M_x$ is
\begin{equation}\label{Mx solution}
    M_x(t, \mathbf{x}) \simeq  g_{aN} M_0 T_2 |\mathcal{A}(\mathbf{x})| \big(1 - e^{-t/T_2}\big) \sin\big(\omega_0 t + \varphi(\mathbf{x})\big),
\end{equation}
where $M_0=n_N\gamma/2$ is the initial magnetization along the $z$-axis for a hyperpolarized spin-1/2 sample, $T_2$ is the transverse relaxation time, $\mathcal{A}(\mathbf{x}) \equiv \pm\partial_x a(\mathbf{x}) + i \partial_y a(\mathbf{x})$ ($+$ for $\gamma>0$, $-$ for $\gamma<0$) is the combination of the derivatives of the axion field that drives $M_x$, $\varphi(\mathbf{x})$ is the phase of $\mathcal{A}(\mathbf{x})$, and we have taken the real part of $M_x$ as the physical solution. 

In the short-time limit $t \ll T_2$, the amplitude of the transverse magnetization grows linearly in time since $1-e^{-t/T_2} \simeq t/T_2$,
while for $t \gg T_2$, the growth saturates and $M_x(t)$ becomes monochromatic:
\begin{equation}\label{Mx t>T2}
    M_x(t, \mathbf{x}) \simeq  g_{aN} M_0 T_2 |\mathcal{A}(\mathbf{x})| \sin\big(\omega_0 t + \varphi(\mathbf{x})\big).
\end{equation}
We analyze this saturated signal in the frequency domain. 
Following the definition and procedure in Ref.~\cite{Dror:2022xpi}, we calculate the power spectrum density (PSD) of $M_x$ in the $t\gg T_2$ limit, for any location (see Supplemental Material for details):
\begin{equation}\label{Pak}
    P^a_k (\mathbf{x}) \simeq \frac{g_{aN}^2 M_0^2 T_2^2 |\mathcal{A}(\mathbf{x})|^2}{\Delta \omega_k^2 T} \, \sin^2\left(\frac{\Delta \omega_k T}{2}\right),
\end{equation}
where $T$ is the observation time, the subscript $k$ corresponds to the frequency index with frequency $\omega_k\equiv2\pi k/T$, $\Delta \omega_k \equiv \omega_k - \omega_0$ is the frequency difference, and we only retain the term that peaks at $k>0$.
The monochromatic signal lies dominantly in a single bin at $k_*=\omega_0T/2\pi$ where the PSD peaks.

\vspace{0.2cm}
\noindent{\bf Experimental Sensitivity.}\label{sec:sensitivity}
For illustration, we consider a double-cylindrical NMR setup consisting of a spin source and a spin sensor with the same geometry, as shown in Fig.~\ref{fig:sensitivity} ({\bf Left}).
To match the Larmor frequencies of nuclear spins in both devices, the source and sensor media are made of the same material and are polarized in the same static magnetic field $\mathbf{B}_0$ aligned with the cylinders’ longitudinal axis ($z$-axis)~\footnote{The $Q$-factor of NMR for a polarized sample with $T_2=100\mathrm{s}$ and Larmor frequency $\omega_0\sim\mathrm{GHz}$ is $Q\equiv\omega_0T_2/2\sim10^9$, resulting in a resonance width $\Delta\omega/\omega_0=Q^{-1}\sim10^{-9}$. The longitudinal field $\mathbf{B}_0$ for the source and the sensor thus needs to be matched with a relative precision of $10^{-9}$ to satisfy the resonance condition for the sensor medium. Current superconducting magnets with $\mathcal{O}(10)\mathrm{T}$ magnetic field can achieve $10^{-9}$ uniformity within centimeter-scale volume with $10^{-8}/\mathrm{hour}$ shift~\cite{wikus2022commercial}, and $10^{-6}$ uniformity within decimeter-scale volume with $10^{-9}/\mathrm{hour}$ shift~\cite{boulant2024vivo}. Persistent efforts are being made to develop more GHz-class NMR magnets with field inhomogeneity below $10^{-9}$ to resolve the narrow spectral lines in structure biology~\cite{wikus2022commercial}. Operating at a lower $\omega_0$ can relax the requirement on field homogeneity. As we will discuss in the next section, reducing $\omega_0$ does not degrade our projected sensitivity.}.
A transverse oscillating magnetic field $\mathbf{B}_{1}(t)$ is applied to drive nuclear spin precession inside the spin source which is enclosed within a magnetic shield that prevents $\mathbf{B}_1(t)$ from affecting the spin sensor (see Supplemental Materials for discussions on the effectiveness of the shielding)~\footnote{The $\mathbf{B}_0$ field applied to the spin source can magnetize the shielding material which induces a residual field, altering the total longitudinal field. This effect is negligible as long as the ratio of the residual field to $B_0$ remains below $Q^{-1}$. Otherwise, it can be solved by introducing an identical magnetic shield for the spin sensor or tuning the longitudinal field with a relative precision of $\mathcal{O}(Q^{-1})$.}, and is appropriately modulated to keep a stably sustained coherent spin precession and we assume the spins to keep hyperpolarized in the $xy$ plane~\footnote{Prolonged spin precession has be achieved for $\mathcal{O}(10^4)$s in $^{129}$Xe and $^3$He gases using spin masers, which utilize optical pumping along the $z$ direction and a rotating transverse magnetic field with a 90\degree phase delay (compared to the spin precession) generated by a feedback system~\cite{bear2000limit,rosenberry2001atomic,yoshimi2002nuclear,yoshimi2012low}. By appropriately choosing the pumping and feedback parameters, the stabilized transverse magnetization can achieve $\mathcal{O}(1)$ the initial magnetization $M_0$~\cite{feng2025nonlinear}. We hope in the future this technique can be further applied to $^{129}$Xe or $^3$He liquids.}.  
The spin sensor is placed adjacent to the spin source (along $x$-axis) to maximize the axion-induced signal. 
The source and sensor could be either within a single solenoid or one could employ two individual solenoids tuned to the same Larmor frequency.

The transverse nuclear magnetization $M_x$ in the sensor medium driven by the axion-induced magnetic field $\mathbf{B}_a$ is measured with a SQUID magnetometer which probes the magnetic field induced by $M_x$ at its pickup coil. Since the axion field declines with the distance from the spin source, the transverse magnetization $M_x$ in the detection sample and its induced magnetic field are not uniform along the $x$ direction. To optimize signal strength, we consider a SQUID pickup coil at the near end of the sample, as depicted in Fig.~\ref{fig:sensitivity} ({\bf Left}) (see Supplemental Material for details). 

To present the prospective detection sensitivity to $g_{aN}$, we choose the source and the detection media to be pure $^{129}\mathrm{Xe}$, which has a number density of $n_N=1.3\times10^{22}\,{\rm cm}^{-3}$ and a nuclear gyromagnetic ratio of $\gamma=-7.40\times 10^{7}\,\mathrm{s^{-1}T^{-1}}$, and we assume the transverse relaxation time of the detection sample to be $T_2 = 100\,\mathrm{s}$ (as chosen in other proposed axion spin-precession experiments \cite{Budker:2013hfa,Dror:2022xpi}). We take $B_0 = 10\,\mathrm{T}$, giving a Larmor frequency $\omega_0 = 0.74\,\mathrm{GHz}$ for $^{129}\mathrm{Xe}$ nuclei, and assume that both the source and detection samples are hyperpolarized.

The results are shown in Fig.~\ref{fig:sensitivity} ({\bf Right}), where we present the projected sensitivities for a pair of cylindrical NMR devices with radius $R=0.1,\, 1,\,5,\,10\,{\rm cm}$, height $H=3R$, and their cylinder axes separated by $3R$. The  95\% C.L. upper limits are derived by the Bayesian likelihood analysis (see Supplemental Material for details). 
For small axion masses ($m_a<\omega_0$), the experiment is in the near-field regime of the produced axion field~\footnote{For the $R=10\,\mathrm{cm}$ setup, the axion field in the spin sensor slightly deviates from the near-field result for small masses.}.
With a 15-day integration time, a pair of $R=1\,{\rm cm}$ NMR devices can achieve sensitivity below axion-nucleon couplings $g_{aN}$ constrained from cooling of neutron stars and observations of supernova 1987A~\footnote{Note that the actual magnetic field that fluxes through a pickup loop near end of the sensor medium is $\mathcal{O}(1)$ smaller than $M_x$ (see Supplemental Material for details). Since $M_x \propto g_{aN}^2$, our sensitivities obtained by treating them as equal is slightly overestimated by an $\mathcal{O}(1)$ factor.}. A similar conclusion can be reached with only a 1-hour integration time and a pair of $R=10\,\mathrm{cm}$ devices.

Since the produced axion field is coherent and oscillates at the Larmor frequency $\omega_0$, independently of its mass, one can achieve a broad-band axion search with a single driving frequency and run the experiment for long durations. Thus, we also project the sensitivities for a 1-year integration time with the aforementioned setups, which are greatly improved.

The maximal axion mass that can be probed with this configuration is set by the inverse radius of the spin source, $m_a \sim 1/R$, above which the exponential suppression in Eq.~\eqref{eq:ax} becomes significant. Therefore, by decreasing the radius of the cylinder, searches can be readily extended to larger axion masses. With $1/R \simeq 20\,\mu\mathrm{eV} (1~\mathrm{cm}/R)$, one can probe axions with masses above the maximal NMR frequency that can be achieved experimentally ($\sim 1\,\mu\mathrm{eV}$), which sets the maximal detectable axion mass for CASPEr\,\footnote{In the $m_a\gg1/R$ regime, the magnetization within the detection sample decreases dramatically and the signal is only contributed by a small portion of the sample around the near end. In Fig.~\ref{fig:sensitivity} ({\bf Right}), we always take $V$ in $P_k^{\mathrm{SP}}$ as the volume of the entire detection sample such that the spin projection noise is underestimated in this regime. The actual sensitivities are thus decaying more rapidly when $m_a \gg 1/R$ than in Fig.~\ref{fig:sensitivity} ({\bf Right}).}.

The sensitivity scaling of the experiment with various parameters can be understood using the signal-to-noise ratio (SNR). The signal lies predominantly in the frequency bin of $k_*=\omega_0 T/2\pi$, such that ${\rm SNR} = P^a_{k_*}/B_{k_*}$, with $B_k=P_k^{\mathrm{SQ}}+P_k^{\mathrm{SP}}$ the background noise PSD, where  $P_k^{\mathrm{SQ}}$ is the SQUID noise and $P_k^{\mathrm{SP}}$ is the spin projection noise~\footnote{This still holds true with the Bayesian framework we adopt, even though our signal is deterministic rather than stochastic. We provide a detailed discussion of noise sources and the likelihood analysis in Supplemental Material.}. 
Setting the SNR to a constant threshold and plugging in the expressions for $P_{k_*}^a$ and $B_{k_*}$ (which is dominated by $P_{k*}^{\mathrm{SP}}$ for the benchmark parameters we consider), we can solve for the scaling behavior of the experimental sensitivity to $g_{aN}$:
\begin{equation}\label{sensitivity scaling}
    g_{aN} \propto n_N^{-3/4} T_2^{-1/4} T^{-1/4} V^{-1/4} |\mathcal{G}(\mathbf{x}_0)|^{-1/2},
\end{equation}
where $V$ is the volume of the detection sample and $\mathcal{G}(\mathbf{x})\equiv\mathcal{A}(\mathbf{x})/g_{aN}n_N$ is a dimensionless function that depends on $\omega_0$, $m_a$ and the geometry of the source sample, $H/R$.
In the configuration we consider, $\mathcal{G}(\mathbf{x})$ is evaluated at the near end of the spin sensor, $\mathbf{x}_0=(2R,0,0)$, with the origin at the center of the source cylinder.

The sensitivity scales simply with $n_N$, $T_2$, $T$, and $V$, which account for the scaling of curves in Fig.~\ref{fig:sensitivity} for different choices of device sizes and interrogation times, while its dependence on $\omega_0$, $m_a$, and geometry is implicitly contained by $|\mathcal{G}(\mathbf{x}_0)|$.
For example, in the near-field limit, with $\mathbf{x}_0=(2R,0,0)$, $|\mathcal{G}(\mathbf{x}_0)|$ attains its maximum at $H/R \simeq 3$, motivating the benchmark source–sensor geometry used for the projections.  
We provide detailed discussions on the general behavior of $|\mathcal{G}(\mathbf{x}_0)|$ in Supplemental Material.

Specific to the 95\% C.L. upper limits of $g_{aN}$ derived with a Bayesian likelihood framework in Fig.~\ref{fig:sensitivity} ({\bf Right}), we take the near-field value, $|\mathcal{G}(\mathbf{x}_0)|\simeq0.05$, for the optimized height-to-radius ratio $H/R=3$, which corresponds to the small $m_a$ region for the aforementioned configurations, then
\begin{align}
    &g_{aN}^{(\mathrm{UL})}\simeq 2\times 10^{-11}\,{\rm GeV}^{-1}\left(\frac{n_N}{1.3 \times 10^{22}\,{\rm cm^{-3}}}\right)^{-3/4}\nonumber\\
    &\times\left(\frac{T_2}{100\,{\rm s}}\right)^{-1/4}\left(\frac{T}{1\,{\rm yr}}\right)^{-1/4}\left(\frac{V}{0.01\,{\rm m^{3}}}\right)^{-1/4},
\end{align} 
where $V\simeq0.01{\rm m}^3$ corresponds to the $R=10\,\mathrm{cm}$ configuration. 

\vspace{0.2cm}
\noindent{\bf Discussion.}
The sensitivity reach can be further improved in future experiments by leveraging an NMR sample material with a larger nucleus density $n_N$ and a longer transverse relaxation time $T_2$, interrogating for an integration time $T$, or building a spin sensor with a larger effective volume $V$ that is efficiently transversely magnetized by the axion field. 
Since the axion field produced by the spin source declines as a power law of $x/R$, the maximal effective volume $V$ that can be achieved should be about the same order of magnitude as the volume of the spin source.
In addition, the size of the effective volume $\sim V^{1/3}$ is also limited by the wavelength of the produced axion field $\sim1/k_0$, beyond which $M_x$ is out of phase and has no positive contribution to the signal (see Supplemental Material for discussions). 

While we adopt $\omega_0\sim \mathrm{GHz}$ --- about the highest NMR frequencies achieved experimentally --- to obtain the sensitivities in Fig.~\ref{fig:sensitivity} ({\bf Right}), a lower $\omega_0$ would neither weaken the sensitivity reach nor reduce the maximal axion mass accessible to the setups. As long as $\omega_0R\ll1$, the low-mass sensitivities are governed by the near-field value, $|\mathcal{G}(\mathbf{x}_0)|\simeq0.05$, and the upper mass limit is set by $m_a\sim 1/R$, both independent of $\omega_0$.
However, if, with either a larger Larmor frequency $\omega_0$ or a larger size of samples $R$ (which requires a precisely uniform static magnetic field of larger size in larger volume), future spin source setups could achieve $\omega_0R\simeq2.5$ where $|\mathcal{G}(\mathbf{x}_0)|$ (for $H/R=3$) reaches its peak value of $\simeq 0.8$, sensitivities in the small $m_a$ region can be strengthened by a factor of 4 compared with the near-field ones.
On the other hand, a larger $\omega_0$ (and hence smaller radius) can extend the sensitivity reach to larger axion masses, since the exponential decay occurs only when $m_a$ exceeds $\omega_0$ and $\kappa_0=\sqrt{m_a^2-\omega_0^2}$ exceeds $1/R$.

Although the experiment can, in principle, be run continuously for a total observation time due to the coherent nature of the produced axion field, as we show in Supplemental Material, the sensitivity of a continuous experiment for an observation time $T$ is equivalent to that of multiple separated experiments with a total integration time $T$. This makes the experiment more applicable and reliable since one can calibrate the apparatus daily or hourly and conduct multiple experiments simultaneously with multiple sets of devices, without weakening the ultimate detection sensitivity.

In summary, we solved for the full dynamical picture of axion production and detection with nuclear spin-precession and found that by combining two NMR experiments, one can achieve considerable sensitivity to axions solely through the axion-nucleon coupling. It is also worth noting that combining NMR devices with resonant cavities or other techniques could allow for the production and detection of axions through both the axion-nucleon and axion-photon couplings, thereby enabling probes of the product of these coupling strengths with possibly better sensitivities. 

\section{Acknowledgments}
The authors thank Russell Bowers,  Peter Graham, Yoonseok Lee, Amalia Madden, and Surjeet Rajendran for useful discussions. The research of JD is supported in part by the U.S. Department of Energy grant number DESC0025569. JD would like to thank the Aspen Center for Physics for hospitality during the completion of part of this work. 

\bibliography{ref}

\clearpage

\onecolumngrid

\newpage

\widetext
 \begin{center}
   \textbf{\large SUPPLEMENTAL MATERIAL \\[.2cm] ``Axion Production and Detection Using a Dual NMR-type Experiment''}\\[.2cm]
  \vspace{0.05in}
  {Jeff A. Dror, Qiushi Wei, and Fengwei Yang}
\end{center}

\setcounter{equation}{0}
\setcounter{figure}{0}
\setcounter{table}{0}
\setcounter{page}{1}
\setcounter{section}{0} 
\makeatletter
\renewcommand{\thesection}{S-\Roman{section}}
\renewcommand{\theequation}{S-\arabic{equation}}
\renewcommand{\thefigure}{S-\arabic{figure}}

\section{Effective axion-nucleus coupling}

Coupled to nuclear spins, the effective coupling constant, $g_{aN}$, is related to the axion-neutron coupling, $g_{an}$, and the axion-proton coupling, $g_{ap}$, through form factors. In the non-relativistic limit, operators $g_{aN}\mathbf{I}$ and $g_{an}\mathbf{S}_{n}+g_{ap}\mathbf{S}_p$ should be equivalent in zero-momentum transfer processes, where $\mathbf{I}$ is the nucleus spin and $\mathbf{S}_n$ and $\mathbf{S}_p$ are the total neutron and proton spins in the nucleus. Specifically,
\begin{align}
    g_{aN}\left< \mathbf{I} \right> = g_{an}\left< \mathbf{S}_n \right> + g_{ap}\left< \mathbf{S}_p \right>, \quad \left< ... \right> \equiv \left< i,i_z|...| i,i_z\right>.
\end{align}
For the $i_z=i$ state, the $z$ component of this equation gives
\begin{align}
    g_{aN} = g_{an}\frac{\left< S_n^z \right>}{i} + g_{ap}\frac{\left< S_p^z \right>}{i}.
\end{align}
For $^{129}$Xe, $i=1/2$, $\left< S_n^z \right>\simeq0.300$, $\left< S_p^z \right>\simeq0.023$~\cite{Stadnik:2014xja,Hu:2021awl}. Thus, $g_{aN} \simeq 0.600g_{an}+0.046g_{ap}$.

\section{Axion production solution}

The equation of motion of axions, Eq.~\eqref{EOM}, can be solved using the retarded Green's function:
\begin{align}
    (\partial^2+m_a^2)G(x;x') = \delta^4(x-x'),~~~ 
    G(x;x') = - \int \frac{d^4k}{(2\pi)^4} e^{-ik\cdot(x-x')} \frac{1}{(\omega-\omega_\mathbf{k}+i\epsilon)(\omega+\omega_\mathbf{k}+i\epsilon)},
\end{align}
where we adopt the metric convention $\eta_{\mu\nu}=(+,-,-,-)$, $\omega_\mathbf{k} = \sqrt{|\mathbf{k}|^2+m_a^2}$, and $\epsilon$ is an infinitesimal positive quantity.
The solution for the produced axion field is thus given by
\begin{align}
\begin{aligned}
    a(x) = \int d^4x'\, G(x;x') j(x'),
\end{aligned}
\end{align}
with $j(x)\equiv-g_{aN}\partial_\mu j^{5\mu}(x)$ the source term.
Adopting the complex representation and factoring out the monochromatic time dependence, i.e., $a(x) = e^{-i\omega_0 t} a(\mathbf{x})$ and $j(x) = e^{-i\omega_0 t} j(\mathbf{x})$, the solution reduces to
\begin{align}
\begin{aligned}
    a(\mathbf{x}) &= -\int \frac{d^3k}{(2\pi)^3}\,e^{i\mathbf{k}\cdot\mathbf{x}}\frac{1}{(\omega_0-\omega_\mathbf{k}+i\epsilon)(\omega_0+\omega_\mathbf{k}+i\epsilon)} \int d^3x'\, e^{-i\mathbf{k}\cdot\mathbf{x'}} j(\mathbf{x}') \\
    &= \int \frac{d^3k}{(2\pi)^3}\,e^{i\mathbf{k}\cdot\mathbf{x}}\frac{\widetilde{j}(\mathbf{k})}{|\mathbf{k}|^2+m_a^2-\omega_0^2-2i\epsilon \omega_0} \\
    &= \int \frac{d^3k}{(2\pi)^3}\,e^{i\mathbf{k}\cdot\mathbf{x}}\frac{\widetilde{j}(\mathbf{k})}{|\mathbf{k}|^2+m_a^2-\omega_0^2-i\epsilon} ~~~ (\text{assuming $\omega_0>0$}),
\end{aligned}
\end{align}
where $\widetilde{j}(\mathbf{k})$ is the Fourier transform of the $j(\mathbf{x})$ and we redefine $\epsilon$ in the last equality.

This integral can be transformed into the position space, recognizing the inverse Fourier transform of $\widetilde{V}({\bf k})=(|{\bf k}|^2+m_a^2-\omega_0^2-i\epsilon)^{-1}$ by closing an appropriate contour for $|\mathbf{k}|$ on the complex plane and using the residue theorem:
\begin{equation}
    V({\bf x})=
    \left\{
    \begin{matrix} 
    \frac{1}{4\pi|{\bf x}|}e^{i\sqrt{\omega_0^2-m_a^2}|{\bf x}|}, &\omega_0\ge m_a\\
    \frac{1}{4\pi|{\bf x}|}e^{-\sqrt{m_a^2-\omega_0^2}|{\bf x}|}, &\omega_0< m_a
    \end{matrix}
    \right..
\end{equation}
Then the axion solution, using the convolution theorem, is transformed into
\begin{equation}
   a({\bf x}) =\int d^3x'\, V({\bf x}-{\bf x'})j({\bf x'}) = \int d^3x'\, \frac{1}{4\pi|\mathbf{x}-\mathbf{x}'|} e^{ik_0 |\mathbf{x}-\mathbf{x}'|} j({\bf x'}),
\end{equation}
and we define $k_0\equiv\sqrt{\omega_0^2-m_a^2}$ (for $\omega_0\geq m_a$) or $i\kappa_0$ with $\kappa_0 \equiv \sqrt{m_a^2-\omega_0^2}$ (for $\omega_0<m_a$).

For a non-relativistic medium with spin-1/2 nuclei, $j^{5\mu}(t,\mathbf{x})\simeq(0,2\mathbf{s}(t,\mathbf{x}))$, where $\mathbf{s}(t,\mathbf{x})$ the nuclear spin density.
Writing $\mathbf{s}(t,\mathbf{x})=e^{-i\omega_0 t}\mathbf{s}(\mathbf{x})$ in the complex representation, the produced axion field is
\begin{align}\label{a solution}
\begin{aligned}
   a({\bf x}) &= -2g_{aN} \int d^3x'\, \frac{1}{4\pi|\mathbf{x}-\mathbf{x}'|} e^{ik_0 |\mathbf{x}-\mathbf{x}'|}\ \nabla'\cdot \mathbf{s}(\mathbf{x}),
\end{aligned}
\end{align}
where $\nabla'$ is the gradient with respect to $\mathbf{x}'$. 
Furthermore, if the source medium is uniform and the nuclear spins are processing in alignment as considered in the main text, then $\mathbf{s}(\mathbf{x})=n_N\boldsymbol{\sigma}/2$ inside the source, where $n_N$ is the nucleus density and $\boldsymbol{\sigma}$ is the complex representation of the nuclear spin unit vector with time dependence factored out, and $\nabla\cdot\mathbf{s}(\mathbf{x})$ is non-vanishing only at the boundary of the source medium.
The solution, Eq.~\eqref{a solution}, is thus transformed into an integral over the closed surface of the source
\begin{align}
    a({\bf x}) &= g_{aN}n_N \oint d^2x'\, \frac{1}{4\pi|\mathbf{x}-\mathbf{x}'|} e^{ik_0 |\mathbf{x}-\mathbf{x}'|}\ \hat{\bf n}'\cdot \boldsymbol{\sigma},
\end{align}
where $\hat{\bf n}'$ denotes the unit normal vector to the source surface at $\mathbf{x}'$.
This is exactly the full-form solution in the main text, Eq.~\eqref{eq:ax}.

\section{The ``spin charge'' picture}

The axion field solution, Eq.~\eqref{a solution}, indicate that the axion field can be interpreted as sourced by a ``spin charge'' distribution induced by the spin polarization in a medium, which is analogous to the electric bound charge induced by the electric polarization in dielectric
\begin{align}
    \begin{aligned}
        \boldsymbol{\sigma}(t) &\longleftrightarrow \mathbf{p}(t,\mathbf{x}) \\
        2\mathbf{s}(t,\mathbf{x}) &\longleftrightarrow \mathbf{P}(t,\mathbf{x})\\
        \rho_\mathrm{s} (t,\mathbf{x})\equiv-2\nabla\cdot\mathbf{s}(t,\mathbf{x}) &\longleftrightarrow \rho_\mathrm{p}(t,\mathbf{x}) = -\nabla\cdot \mathbf{P}(t,\mathbf{x})\\
        \Sigma_\mathrm{s} (t,\mathbf{x})\equiv 2\hat{\bf n}\cdot\mathbf{s}(t,\mathbf{x}) &\longleftrightarrow \Sigma_\mathrm{p}(t,\mathbf{x}) = \hat{\bf n}\cdot \mathbf{P}(t,\mathbf{x})\\
    \end{aligned}
    ~~~
    \begin{aligned}
        &(\text{Electric dipole})\\
        &(\text{Electric polarization})\\
        &(\text{Bound charge density})\\
        &(\text{Bound charge surface density})\\
    \end{aligned}
\end{align}
The axion field solution, after adding back the harmonic time dependence, is therefore written in terms of ``spin charge'' as
\begin{align}\label{a from spin charge}
    a(t,{\bf x}) = g_{aN} \int d^3x'\, \frac{1}{4\pi|\mathbf{x}-\mathbf{x}'|} e^{ik_0 |\mathbf{x}-\mathbf{x}'|}\,\rho_\mathrm{s}(t,\mathbf{x}'),
\end{align}
where the integration is performed over the entire space such that the contribution of surface charge is already included.

In the near-field limit, where $k_0|{\bf x}-{\bf x'}|\ll1$ (for $\omega_0\geq m_a$) or $\kappa_0|{\bf x}-{\bf x'}|\ll1$ (for $\omega_0<m_a$), the produced axion field and it gradient reduce to simple forms
\begin{align}
    a(t,{\bf x}) = g_{aN} \int d^3x'\, \frac{1}{4\pi|\mathbf{x}-\mathbf{x}'|} \rho_\mathrm{s}(t,\mathbf{x}'),~~~~ \nabla a(t,{\bf x}) = -g_{aN} \int d^3x'\, \frac{\mathbf{x}-\mathbf{x}'}{4\pi|\mathbf{x}-\mathbf{x}'|^3} \rho_\mathrm{s}(t,\mathbf{x}').
\end{align}
These equations are exactly in the same form as those of electrostatics if we draw an analogy
\begin{align}
    \begin{aligned}
        g_{aN} &\longleftrightarrow 1/\epsilon_0\\
        a(t,\mathbf{x}) &\longleftrightarrow \Phi\\
        \nabla a(t,\mathbf{x}) &\longleftrightarrow -\mathbf{E}
    \end{aligned}
    ~~~
    \begin{aligned}
        &(\text{Vacuum permittivity})\\
        &(\text{Electric potential})\\
        &(\text{Electric field})
    \end{aligned}
\end{align}
Therefore, for any moment $t$, the near-zone axion field $a(t,\mathbf{x})$ and its gradient $\nabla a(t,\mathbf{x})$ has the same configuration as $\Phi$ and $-\mathbf{E}$ produced by a static electric charge given by $\rho_{\mathrm{s}}(t,\mathbf{x})$ at this moment.

\begin{figure*}[!htbp]
        \centering
         \includegraphics[width=0.75\linewidth]{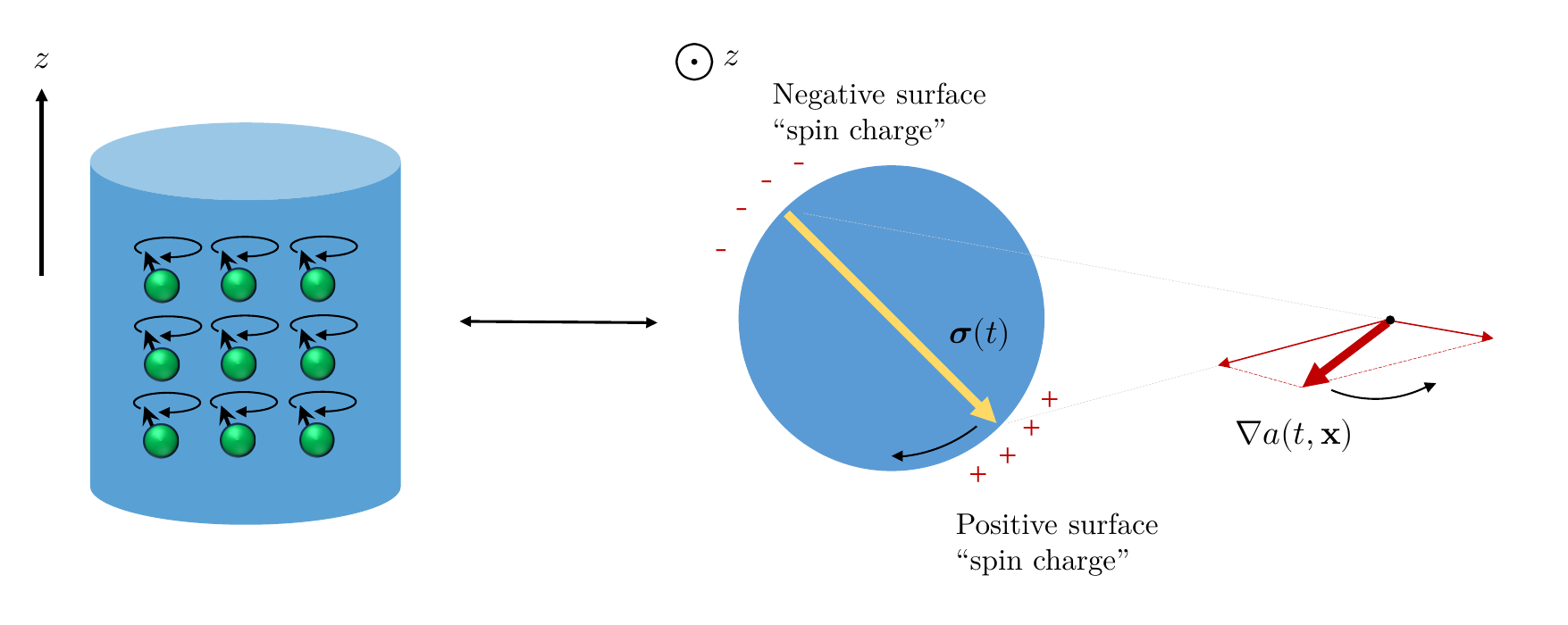}        
    \caption{The ``spin charge'' distribution in a cylindrical, uniform, spin precessing medium and the gradient of axion field it produces in the near zone.}
    \label{fig:spin charge}
\end{figure*}

This method is useful for inferring the behavior of $\nabla a(t,\mathbf{x})$ in the near zone without solving the equations or carrying out the integrals. 
For the ``spin charge'' distribution at any instant $t$, positive ``spin charge'' contributes to a component of $\nabla a(t,\mathbf{x})$ that points towards it, while negative ``spin charge'' contributes a component of $\nabla a(t,\mathbf{x})$ that points away from it.
In Fig.~\ref{fig:spin charge}, we provide a clearer illustration of the ``spin charge'' distribution in a cylindrical, uniform, spin-precessing medium and the produced $\nabla a(t,\mathbf{x})$ in the near zone.
The combined total $\nabla a(t,\mathbf{x})$ rotates in the opposite direction to the nuclear spins $\boldsymbol{\sigma}(t)$.

\section{Bloch equation under the axion-induced magnetic field}

In this section, we solve the Bloch equation for the detection medium discussed in the main text under the axion-induced magnetic field $\mathbf{B}_a$ and derive the transverse magnetization Eq.~\eqref{Mx solution}. The Bloch equation is
\begin{equation}
    \frac{d \mathbf{M}}{dt} = \mathbf{M} \times \gamma \mathbf{B} 
    - \frac{M_x \hat{\mathbf{x}} + M_y \hat{\mathbf{y}}}{T_2} 
    - \frac{(M_z - M_0) \hat{\mathbf{z}}}{T_1},
\end{equation}
where $M_0 \hat{\mathbf{z}}$ is the initial magnetization, $T_1$ and $T_2$ are the longitudinal and transverse relaxation times, and all quantities are evaluated at the same location. 

We decompose the magnetic field as $\mathbf{B} = B_0 \hat{\mathbf{z}} + \mathbf{B}_a$.
Since $|\mathbf{B}_a| \ll B_0$, it is sufficient to work perturbatively in the axion field, treating the magnetization in the $x$ and $y$ direction as first-order corrections. The transverse components obey
\begin{align}
\begin{aligned}
    \frac{d M_x}{dt} &= M_y \gamma B_0 - M_0 \gamma B_{ay} - \frac{M_x}{T_2},\\
    \frac{d M_y}{dt} &= - M_x \gamma B_0 + M_0 \gamma B_{ax} - \frac{M_y}{T_2}.
\end{aligned}
\end{align}
These coupled first-order differential equations can be decoupled by taking second derivatives. 
The equations take a slightly different form depending on the sign of $\gamma$. For simplicity, we assume $\gamma>0$ in what follows, in which case $M_x$ satisfies
\begin{equation}\label{Mx equation}
    \ddot{M}_x + \frac{2}{T_2} \dot{M}_x + \omega_0^2 M_x \simeq \gamma M_0 (\omega_0 B_{ax} - \dot{B}_{ay}),
\end{equation}
where we replace $|\gamma B_0|$ with the Larmor frequency $\omega_0$ and terms of $\mathcal{O}(1/\omega_0 T_2)$ are neglected. Eq.~\eqref{Mx equation} describes a damped harmonic oscillator driven by a periodic force $\mathcal{F}(t,{\bf x}) \equiv\gamma M_0 (\omega_0 B_{ax} - \dot{B}_{ay})$. Assuming the axion field is turned on at $t=0$, given initial conditions $M_x(0) = M_y(0) = 0$, the solution is
\begin{equation}
\begin{aligned} \label{Mx green's function}
    M_x(t,\mathbf{x}) = \frac{1}{\omega_0} \int_0^t dt' \, e^{-(t-t')/T_2} \sin\big(\omega_0 (t-t')\big) \mathcal{F}(t',\mathbf{x}).
\end{aligned}
\end{equation}

If the detection medium is on resonance --- the Larmor frequency of the sample matches the frequency of the axion field --- then by substituting the axion-induced magnetic field form Eq.~\eqref{B_a} and factoring out the harmonic time dependence, the driving force becomes 
\begin{equation}\label{driving force}
    \mathcal{F}(t,\mathbf{x}) = 2 g_{aN} M_0 \omega_0 \mathcal{A}(\mathbf{x}) e^{-i \omega_0 t},
\end{equation}
with $\mathcal{A}(\mathbf{x}) \equiv \partial_x a(\mathbf{x}) + i \partial_y a(\mathbf{x})$.
Carrying out the integral in Eq.~\eqref{Mx green's function} yields the transverse magnetization 
\begin{equation}
    M_x(t, \mathbf{x}) \simeq  g_{aN} M_0 T_2 |\mathcal{A}(\mathbf{x})| \big(1 - e^{-t/T_2}\big) \sin\big(\omega_0 t + \varphi(\mathbf{x})\big),
\end{equation}
where $\varphi(\mathbf{x})$ is the phase of $\mathcal{A}(\mathbf{x})$ and we have taken the real part as the physical solution. While we assumed $\gamma>0$ in its derivation, the same result holds for  $\gamma<0$ with $\mathcal{A}(\mathbf{x}) = -\partial_x a(\mathbf{x}) + i \partial_y a(\mathbf{x})$, in contrast to $\mathcal{A}(\mathbf{x}) = \partial_x a(\mathbf{x}) + i \partial_y a(\mathbf{x})$.

\section{Noise sources\label{sec:noise}}

There are three types of Gaussian background noise in the measurements: thermal noise, SQUID noise, and spin-projection noise. The thermal noise stems from the thermal fluctuations of the readout circuit and can be controlled by cooling the apparatus. We assume it is effectively eliminated. 

The SQUID magnetometer is used to measure the transverse magnetic field induced by the transverse magnetization of the detection sample. 
We treat the transverse magnetic field as equal to the transverse magnetization and uniform within the pickup coil area for the specific setup discussed in the main text. The SQUID noise power spectrum density (PSD) is given by a white noise spectrum~\cite{Anton:2013nrw,Kahn:2016aff,Dror:2022xpi}
\begin{equation}
\label{eq:P_SQ}
    P_k^{\rm SQ}=\frac{1}{A_{\rm eff}^2}P_{\Phi\Phi}^{\rm SQ},
\end{equation}
where $P_{\Phi\Phi}^{\rm SQ}\simeq (10^{-7}\Phi_0)^2/{\rm Hz}$, with $\Phi_0$ the magnetic flux quantum, and $A_{\rm eff}\simeq 0.3\,{\rm cm}^2\left(\frac{A}{80\,{\rm cm}^2}\right)$ the effective sample area sensed by a pickup loop with area $A$, rescaled from the estimation used in Ref.~\cite{Budker:2013hfa}. The resulting SQUID noise considered here for an $R=5$\,cm sample approximately matches that used in Ref.~\cite{Budker:2013hfa}. 

The spin-projection noise stems from the quantum nature of the nuclear spin in the detection medium~\cite{Braun_2007}. 
In the $t\gg T_2$ limit, The PSD of the spin-projection noise is described by a Lorentzian shape~\cite{Dror:2022xpi}
\begin{equation}
\label{eq:P_SP}
    P_{k}^{\rm SP}=\frac{\gamma^2n_NJ}{2V}\frac{T_2}{1+\Delta\omega_k^2T_2^2},
\end{equation}
where $\gamma$ is the gyromagnetic ratio of the nuclei, $n_N$ is the density of nuclei in the detection sample, $J$ is nuclear spin, $V$ is the volume of the detection sample, and we assume the detection sample is hyperpolarized.

In addition to the Gaussian noises, the $\mathbf{B}_1$ field applied to the spin source and an additional transverse RF magnetic field induced by the magnetization of the nuclear spins in the spin source may contaminate the axion-induced magnetic field $\mathbf{B}_a$. This can be avoided by enclosing the spin source in a magnetic shielding that screens transverse RF magnetic fields.
To display technological feasibility, note that the skin depth of copper at room temperature is $\sqrt{2\rho/\omega\mu}=5~\mu {\rm m}(1~{\rm GHz/\omega})^{0.5}$, where $\rho$ and $\mu$ are the resistivity and the relative permeability of the shielding material and $\omega$ is the angular frequency of the oscillating magnetic field. Assuming the contamination field is $|\mathbf{B}_1|\sim \mathrm{\mu T}$ with 1-GHz angular frequency, an 0.5 mm-thick copper shield can already screen it down to $3\times10^{-50} \,\mathrm{T}$, far below any axion-induced magnetic field our spin sensor can probe. 
We assume the reflected magnetic field from the surface of the magnetic shielding can be controlled by absorption. 
One more subtlety about the magnetic shielding is that the axion field may interact with the shielding material if it is spin-polarized under $\mathbf{B}_0$. The resulting transverse magnetization will source an additional oscillating magnetic field which can contaminate the detection. However, by estimation, we find this contaminating magnetic field is sub-dominant compared to the axion-induced magnetic field due to the suppression from deviating from resonance condition, $\omega_0\neq\omega_{0,{\rm s}}$, where $\omega_0$ is the frequency of the axion field and $\omega_{0,{\rm s}}$ is the Larmor frequency of the shielding material.

\section{Likelihood Framework\label{sup:likelihood}}

In this work, we use a likelihood framework in the frequency domain to compute the projected sensitivities.
Suppose an experiment begins at $t=0$ after the spin sensor has been exposed to the axion field for $\tau \gg T_2$, where magnetization $M_x$ is measured on a discrete time series $\{n \Delta t\}$, $n=0,1,\dots,N-1$, over a duration $T = N \Delta t$.  
The discrete Fourier transform (DFT) and the power spectrum density (PSD) of $M_x$ are defined as
\begin{equation}
\begin{aligned}
    \widetilde{M}_k (\mathbf{x}) &\equiv \sum_{n=0}^{N-1} e^{-i 2\pi k n / N} M_x(n\Delta t, \mathbf{x}), \quad  P_k(\mathbf{x}) \equiv \frac{\Delta t^2}{T} |\widetilde{M}_k (\mathbf{x})|^2,
\end{aligned}
\end{equation}
where $k=0,1,...,N-1$.
If the sampling is sufficiently dense so that $\Delta t$ is much smaller than the characteristic timescale of the variation of $M_x$, $\widetilde{M}_k$ can be approximated by the short-time Fourier transform (STFT) on $0 \leq k \leq N/2$ :
\begin{equation}
\begin{aligned}
    \widetilde{M}_k (\mathbf{x}) \simeq \frac{1}{\Delta t} \int_0^T dt \, e^{-i 2\pi k t / T} M_x(t, \mathbf{x}).
\end{aligned}
\end{equation}
For $(N+1)/2 \leq k < N$, the DFT and PSD are mirrored with those on $0 \leq k \leq N/2$, through
\begin{align}\label{eq:mirror}
    \widetilde{M}_k (\mathbf{x}) = \widetilde{M}^*_{N-k} (\mathbf{x}), \quad P_k (\mathbf{x}) = P_{N-k} (\mathbf{x}).
\end{align}

Plugging in the expression of the axion-induced magnetization, Eq.~\eqref{Mx t>T2}, yields, on $0 \leq k \leq N/2$:
\begin{align}\label{M_x DFT}
    \widetilde{M}^a_k (\mathbf{x}) &\simeq -\frac{ig_{aN} M_0 T_2 |\mathcal{A}(\mathbf{x})|}{\Delta\omega_k T/N} \sin \left[\frac{1}{2} \Delta \omega_k T \right]\exp\left[i\left(\phi+\varphi(\mathbf{x})-\frac{1}{2}\Delta \omega_k T\right)\right],\\
    P^a_k(\mathbf{x}) &\simeq \frac{g_{aN}^2 M_0^2 T_2^2 |\mathcal{A}(\mathbf{x})|^2}{\Delta\omega_k^2 T} \sin^2 \left[\frac{1}{2} \Delta \omega_k T \right],
\end{align}
where $\Delta\omega_k \equiv \omega_k-\omega_0$, $\phi \equiv \omega_0 \tau$ which is technically a random phase if $\tau$ is not measured, $\varphi(\mathbf{x})$ is the phase of $\mathcal{A}(\mathbf{x})$, and we have dropped the terms that peak at negative $k$.
Therefore, according to Eq.~\eqref{eq:mirror}, the DFT and PSD of the axion-induced magnetization exhibit two peaks on $0\leq k<N$ at $k_*=\omega_0T/2\pi$ and $N-k_*=N-\omega_0T/2\pi$.
The noise PSD expressions, Eqs.~\eqref{eq:P_SQ} and~\eqref{eq:P_SP}, are similarly derived with STFT and defined on $0\leq k\leq N/2$~\cite{Dror:2022xpi}. 

Supposing there are in total of $r$ such experiments with the experimental setup discussed in the main text, the DFT and the PSD of the axion induced signal for each of the experiments are, 
on $0\leq k \leq N/2$, 
\begin{align}
    \widetilde{M}_k^{a(i)} &= -\frac{ig_{aN} M_0 T_2 |\mathcal{A}(\mathbf{x}_0)|}{\Delta\omega_k T/N} \sin \left[\frac{1}{2} \Delta \omega_k T \right] \exp\left[{i\left(\phi^{(i)}+\varphi(\mathbf{x}_0)-\frac{1}{2}\Delta \omega_k T\right)}\right],\\
    P^{a(i)}_k &= \frac{g_{aN}^2 M_0^2 T_2^2 |\mathcal{A}(\mathbf{x}_0)|^2}{(\Delta\omega_k)^2 T} \sin^2 \left[\frac{1}{2} \Delta \omega_k T \right],
\end{align}
where the superscript $(i)$ denotes $i$-th experiment, $\mathbf{x}_0$ is the near end of the spin sensor to the source where we place the pick-up loops and $\phi^{(i)}$ is the phase gap between the equipments' powering on and the first measurement in the $i$-th experiment.
Since the typical frequency of a NMR experiment is about GHz, we can reasonably treat each $\phi^{(i)}$ as uniformly distributed on $[0,2\pi)$.

The likelihood function is therefore
\begin{equation}
\begin{aligned}
    &\mathcal{L} (\{\widetilde{M}_k^{(i)}\}|m_a,g_{aN},\{B_k\}) = \prod_{i=1}^{r} \prod_{k=0}^{N-1} \int_0^{2\pi} d\phi^{(i)}~ p(\phi^{(i)}) \frac{1}{\pi B_k T/\Delta t^2} \exp \left[ -\frac{|\widetilde{M}_k^{(i)} - \widetilde{M}_k^{a(i)}|^2}{B_k T/\Delta t^2} \right] \\
    &= \prod_{i=1}^{r} \prod_{k=0}^{N-1} \frac{1}{\pi B_k T/\Delta t^2} \exp \left[ -\frac{|\widetilde{M}_k^{(i)}|^2 + |\widetilde{M}_k^{a(i)}|^2}{B_k T/\Delta t^2} \right] \int_0^{2\pi} d\phi^{(i)}~ p(\phi^{(i)}) \exp \left[\frac{\widetilde{M}_k^{(i)*}\widetilde{M}_k^{a(i)} + \widetilde{M}_k^{(i)}\widetilde{M}_k^{a(i)*}}{B_k T/\Delta t^2} \right],
\end{aligned}
\end{equation}
where $B_k = P_k^{\mathrm{SQ}} + P_k^{\mathrm{SP}}$ is the background PSD and $p(\phi^{(i)}) = 1/2\pi$ is the prior of $\phi^{(i)}$. 
As $\widetilde{M}_k^{a(i)} \propto \exp[i\phi^{(i)}]$, the integral in the last expression can be evaluated with the equality
\begin{equation}
    \frac{1}{2\pi} \int_0^{2\pi} d\phi~ \exp\left[\alpha e^{i\phi} + \alpha^* e^{-i\phi}\right] = \frac{1}{2\pi} \int_0^{2\pi} d\phi~ e^{|\alpha| \cos \phi} = I_0 (|\alpha|),
\end{equation}
where $I_0$ is the modified Bessel function of the first kind of order zero.
We thus obtain the expression of the likelihood function:
\begin{equation}
    \mathcal{L} (\{\widetilde{M}_k^{(i)}\}|m_a,g_{aN},\{B_k\}) =\prod_{i=1}^{r} \prod_{k=0}^{N-1} \frac{1}{\pi B_k T/\Delta t^2} I_0 \left( \frac{|\widetilde{M}_k^{(i)}||\widetilde{M}_k^{a(i)}|}{B_k T/\Delta t^2} \right) \exp \left[ -\frac{|\widetilde{M}_k^{(i)}|^2 + |\widetilde{M}_k^{a(i)}|^2}{B_k T/\Delta t^2} \right].
\end{equation}

In order to compute the 95\% C.L. exclusion lines, we apply the likelihood-ratio test statistics and the Asimov dataset~\cite{Cowan:2010js}, taking $|\widetilde{M}_{k,\mathrm{Asimov}}^{(i)}| = \sqrt{B_k T/\Delta t^2}$.
The test statistics is given by
\begin{equation}
    q = -2 \ln\left( \frac{\mathcal{L}(\{\widetilde{M}_{k,\mathrm{Asimov}}^{(i)}\}|m_a,0,\{B_k\})}{\mathcal{L}(\{\widetilde{M}_{k,\mathrm{Asimov}}^{(i)}\}|m_a,g_{aN},\{B_k\})} \right) 
    = - 2r \sum_{k=0}^{N-1} \left[ \frac{P_k^a}{B_k} - \ln I_0\left( \sqrt{\frac{P_k^a}{B_k}} \right) \right],
\end{equation}
where we have dropped the superscript $(i)$ in $P_k^{a(i)}$ since they are the same for all $i$'s.
The 95\% C.L corresponds to setting $q=-2.71$.
Since the function $x^2-\ln I_0(x)$ is positive and monotonically increasing in $x$, the condition $q=-2.71$ indicates $P_k^a/B_k \lesssim 2$ for each $k$.
Thus, making use of the approximation $\ln I_0(x) \simeq x^2/4$, the test statistics can be further simplified as
\begin{equation}\label{q sum}
    q \simeq - \frac{3}{2} r \sum_{k=0}^{N-1} \frac{P_k^a}{B_k}.
\end{equation}

As we have discussed above, $P_k^a$ has two peaks, one at $k_* = \omega_0 T/2\pi$ and the other at $N-k_* = N- \omega_0 T/2\pi$ which are mirror duplicates of each other.
Around each peak, $P_k^a$ scales as $\sin^2 (\pi \Delta k)/\Delta k^2$, with $\Delta k \equiv k-k_*$ or $k-(N-k_*)$.
If $k_*$ is an integer, all $k$'s except for the peaks $k_*$ and $N-k_*$ lie exactly at the zeros of $P_k^a$ and thus the sum in Eq.~\eqref{q sum} is contributed only by the two peaks.
It can be shown that for our background PSD $B_k$ whose peaks span much broader than those of $P_k^a$, the sum can always be approximated by adding up the peak values at $k_*$ and $N-k_*$, whether $k_*$ is an integer or not.
This gives
\begin{equation}
    q \simeq  - \frac{3}{2} r \left( \frac{P_{k_*}^a}{B_{k_*}} + \frac{P_{N-k_*}^a}{B_{N-k_*}} \right) = -\frac{3g_{aN}^2 M_0^2 T_2^2 T_{\mathrm{tot}}|\mathcal{A}(\mathbf{x}_0)|^2}{4B_{k_*}},
\end{equation}
where $T_{\mathrm{tot}} = rT$ is the total observation time of the $r$ experiments.
Therefore, regardless of the number of experiments carried out, the sensitivity only depends on the total integration time, $T_\mathrm{tot}$. 

For our benchmark parameter choices, $P_{k_*}^\mathrm{SP}$ dominates $B_{k_*}$.
Plugging in $P_{k_*}^{\mathrm{SP}} \propto \gamma^2 n_N T_2 /V$ and $M_0 = \gamma n_N/2$ since we assume that the spin sensor is hyperpolarized, we obtain the dependence of the upper limit of $g_{aN}$ on the parameters:
\begin{equation}\label{upper limit scaling}
    g_{aN}^{(\mathrm{UL})} \propto n_N^{-3/4} T_2^{-1/4} T_{\mathrm{tot}}^{-1/4} V^{-1/4} |\mathcal{G}(\mathbf{x}_0)|^{-1/2},
\end{equation}
where $\mathcal{G}(\mathbf{x}) \equiv \mathcal{A}(\mathbf{x})/g_{aN}n_N$ is a dimensionless function, which contains all of the dependence of $g_{aN}^{(\mathrm{UL})}$ on $m_a$, $\omega_0$ and the geometry of the source.

\section{Geometrical Dependence of the Signal\label{sup:geometric}}

In this section, we show in detail the dependence of the signal on the height-to-radius ratio of the cylindrical spin source, $H/R$. 
As discussed in the main text, this geometric dependence is contained by the dimensionless function $\mathcal{G}(\mathbf{x}) \equiv \mathcal{A}(\mathbf{x})/g_{aN}n_N$.
It affects the sensitivity since it determines the oscillating driving force of the spin precession in the sensor medium, Eq.~\eqref{driving force}.

Plugging in the expression of the produced $a(\mathbf{x})$, Eq.~\eqref{a solution}, with $\boldsymbol{\sigma} = \hat{\bf x}\mp i\hat{\bf y}$ and $\mathcal{A}(\mathbf{x}) = \pm \partial_x a(\mathbf{x}) + i \partial_y a(\mathbf{x})$, and integrating by parts, we arrive at a compact expression for $\mathcal{G}(\mathbf{x})$:
\begin{equation}\label{G(x)}
\begin{aligned}
    \mathcal{G}(\mathbf{x}) &= -\left(\pm\partial_x + i \partial_y\right)\int_{\mathrm{Source}} d^3 x' \left(\hat{\bf x}\mp i\hat{\bf y}\right) \cdot \nabla \left( \frac{1}{4\pi|\mathbf{x}-\mathbf{x}'|} e^{ik_0|\mathbf{x}-\mathbf{x}'|} \right) \\
    &\begin{aligned}
    &= \mp \int_{\mathrm{Source}} d^3 x' \left(\partial_x^2 + \partial_y^2\right) \left( \frac{1}{4\pi|\mathbf{x}-\mathbf{x}'|} e^{ik_0|\mathbf{x}-\mathbf{x}'|} \right) \\
    &= \mp \int_{\mathrm{Source}} d^3 \tilde{x}' \left(\partial_{\tilde{x}}^2 + \partial_{\tilde{y}}^2\right) \left( \frac{1}{4\pi|\tilde{\mathbf{x}}-\tilde{\mathbf{x}}'|} e^{ik_0 R|\tilde{\mathbf{x}}-\tilde{\mathbf{x}}'|} \right),
    \end{aligned}
    ~~~ (- \text{ for } \gamma>0,\ + \text{ for } \gamma<0)
\end{aligned}
\end{equation}
where in the second equality we rewrite the expression in terms of dimensionless quantities normalized by the radius of the source $R$, with $\tilde{\mathbf{x}} = \mathbf{x}/R$ and $\tilde{\mathbf{x}}' = \mathbf{x}'/R$.
Therefore, for a cylindrical spin source, $\mathcal{G}({\mathbf{x}})$ only depends on the normalized wavelength $k_0 R$, the normalized evaluation location $\mathbf{x}/R$, and the height-to-radius ratio $H/R$.

We consider the near-field limit which gives the largest signals.
It is satisfied by our signals for low-mass axions ($m_a \lesssim 1/R$) since $\omega_0 R \lesssim 1$ with our parameter choices.
In this limit, the exponential factor in Eq.~\eqref{G(x)} can be dropped and
\begin{equation}\label{G near zone}
\begin{aligned}
    \mathcal{G}(\mathbf{x}) =\mp\int_{\mathrm{Source}} d^3 \tilde{x}' \left(\partial_{\tilde{x}}^2 + \partial_{\tilde{y}}^2\right) \frac{1}{4\pi|\tilde{\mathbf{x}}-\tilde{\mathbf{x}}'|}
\end{aligned}
\end{equation}
is real and only depends on $H/R$ if we fix evaluate location $\mathbf{x}/R$.

\begin{figure*}[!htbp]
        \centering
         \includegraphics[width=0.5\linewidth]{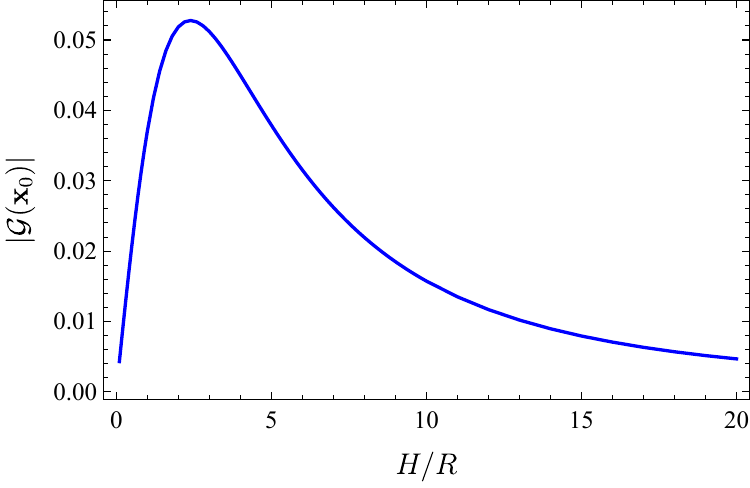}        
    \caption{ The amplitude of the driving force as a function of the height-to-radius ratio. }
    \label{fig:signal_HoverR}
\end{figure*}

Fig.\,(\ref{fig:signal_HoverR}) shows the behavior of $|\mathcal{G}(\mathbf{x})|$ in the near-field limit at $\mathbf{x}_0=(2R,0,0)$, where the near end of the spin sensor is located, by varying $H/R$ from 0.1 to 20. When $H/R\lesssim 0.9$, it grows linearly because a larger $H/R$ gives a larger spin volume for a given $R$ and increases the signal. However, the growth saturates and it decreases by further increasing $H/R$ because the contributions to $\nabla a$ from the portions of the cylindrical source far above or below $x$-axis are opposite to those from the portion near $x$-axis (which can be seen using the ``spin charge'' picture) and there is thus an cancellation in the signal.
The maximum occurs at $H/R \simeq 3$, which justifies our choices of $H/R=3$ in the main text.

\section{Magnetic flux from the detection sample}

In this experiment, the axion-induced NMR signal is detected via the transverse magnetic flux at the SQUID pickup loop produced by the transverse magnetization $M_x$ in the sensor medium and we treat the transverse magnetic field in the pickup coil area as equal to $M_x$ at the near end of the spin sensor on the $x$-axis to derive our result.
In this section, we show that the transverse magnetic field $B_x$ is maximized and related to $M_x$ with an $\mathcal{O}(1)$ factor at the near end of the sensor medium, though our axion induced $M_x$ is not uniform in the sensor medium, different from the case of axion dark matter discussed in the Supplemental Material of Ref.~\cite{Dror:2022xpi}.

The vector potential produced by the magnetization of a medium is
\begin{align}
\begin{aligned}
    \mathbf{A}(t,\mathbf{x}) &= \int d^3x'\, \frac{\nabla'\times \mathbf{M}(t_{\mathrm{ret}},\mathbf{x}')}{4\pi |\mathbf{x}-\mathbf{x}'|},
\end{aligned}
\end{align}
where $t_\mathrm{ret}=t-|\mathbf{x}-\mathbf{x}'|$ is the retarded time.
If the pickup coil is placed in the near zone of the magnetized medium such that $\omega_0 |\mathbf{x}-\mathbf{x}'| \ll 1$, the vector potential reduces to
\begin{align}
\begin{aligned}
    A(t,\mathbf{x}) \simeq \int d^3x'\, \frac{\nabla'\times \mathbf{M}(t,\mathbf{x}')}{4\pi |\mathbf{x}-\mathbf{x}'|}   =\nabla\times \int d^3x'\, \frac{\mathbf{M}(t,\mathbf{x}')}{4\pi |\mathbf{x}-\mathbf{x}'|}.
\end{aligned}
\end{align}
The magnetic field is thus
\begin{align}\label{bx}
\begin{aligned}
    \mathbf{B}(t,\mathbf{x}) &= \nabla\times\mathbf{A}(t,\mathbf{x}) = \nabla\times\left(\nabla\times \int d^3x'\, \frac{\mathbf{M}(t,\mathbf{x}')}{4\pi |\mathbf{x}-\mathbf{x}'|}\right) \\
    &=\nabla\left(\nabla\cdot \int d^3x'\, \frac{\mathbf{M}(t,\mathbf{x}')}{4\pi |\mathbf{x}-\mathbf{x}'|}\right) - \nabla^2\left(\int d^3x'\, \frac{\mathbf{M}(t,\mathbf{x}')}{4\pi |\mathbf{x}-\mathbf{x}'|}\right) \\
    &=\nabla\left(\nabla\cdot \int d^3x'\, \frac{\mathbf{M}(t,\mathbf{x}')}{4\pi |\mathbf{x}-\mathbf{x}'|}\right) + \mathbf{M}(t,\mathbf{x}),
\end{aligned}
\end{align}
where $\mathbf{M}(t,\mathbf{x})=0$ when ${\bf x}$ is outside of the sensor medium.
   
For a cylindrical detection medium as discussed in the main text with its center placed on the $x$-axis, we consider the transverse magnetic field on the $x$-axis $B_x(t,x\hat{\bf x})$
\begin{align}
    B_x(t,x\hat{\bf x}) = \partial_x^2 \int d^3x'\, \frac{M_x(t,\mathbf{x}')}{4\pi |x\hat{\bf x}-\mathbf{x}'|} + M_x(t,x\hat{\bf x}),
\end{align}
where the $\partial_y$ term vanishes and we neglect $M_z$ since its oscillatory component is at higher order.

The spatial dependence of $M_x$ is contained by $\mathcal{G}(\mathbf{x})\equiv \mathcal{A}(\mathbf{x})/g_{aN}n_N$ (Eq.~\eqref{Mx t>T2}).
In particular, in the $m_a\leq \omega_0$ regime, the phase of $M_x$ is determined by the phase of $\mathcal{G}(\mathbf{x})$ which is spatially dependent.
If the size of the detection medium is much smaller compared to the axion wavelength $1/k_0$, $M_x$ is in phase throughout the medium leading to a bulk transverse magnetization and a constructively contributed $B_x$. For a detection medium that spans multiple wavelengths, $M_x$ precesses with different phases at different portions of the medium, which suppresses the production of $B_x$. 
In the $m_a>\omega_0$ regime, $\mathcal{G}(\mathbf{x})$ is real and $M_x$ is in phase throughout the space.
However, if the size of the detection medium exceeds the photon wavelength $1/\omega_0$, the generation of $B_x$ would also undergo some cancellation due to the retardation. 
The determination of these suppression effects requires calculations based on specific experimental configurations.

\begin{figure*}[h!t]
        \centering
         \includegraphics[width=0.465\linewidth]{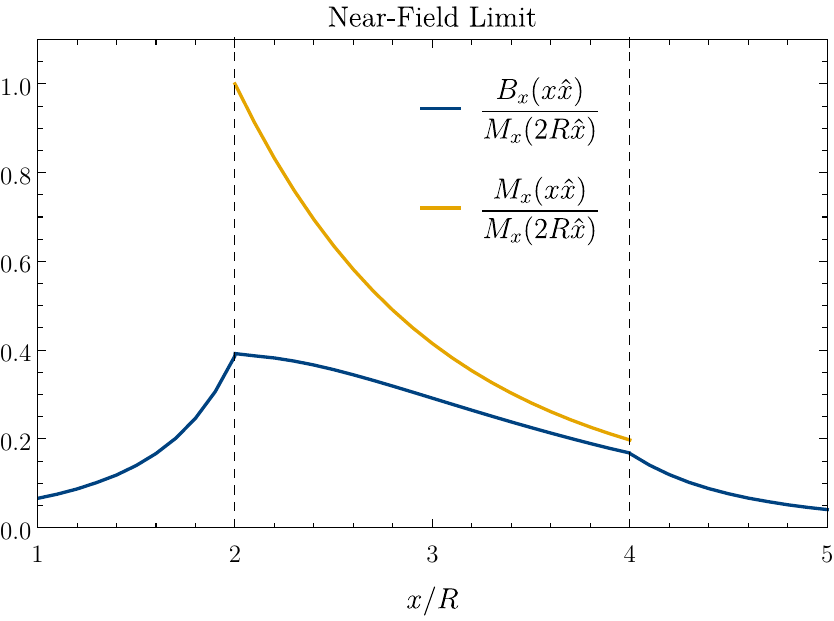}
         \qquad\includegraphics[width=0.475\linewidth]{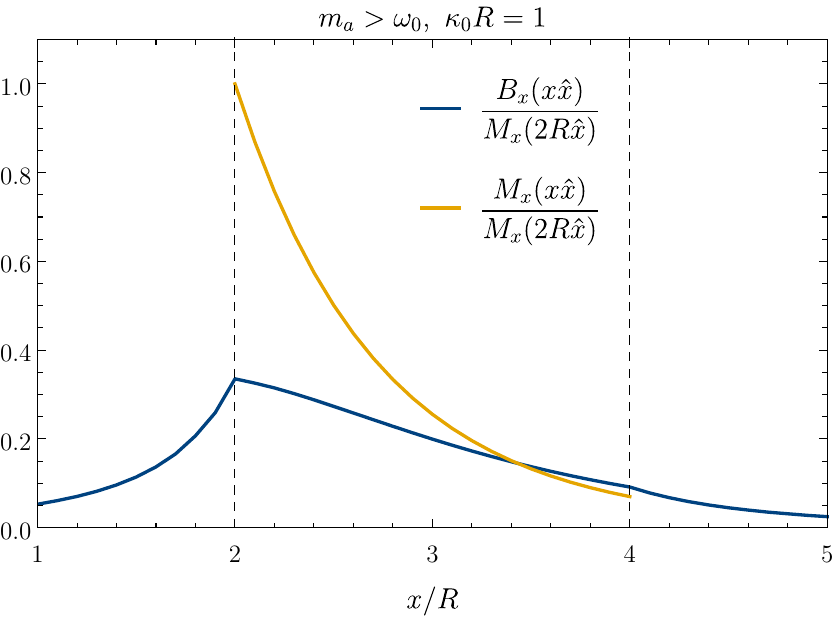}
    \caption{Amplitude of $B_x$ and $M_x$ along $x$-axis for the two-cylinder setups in the main text (height-to-radius ratio $H/R=3$, near end of the sensor at $\mathbf{x}_0=(2R,0,0)$). \textbf{Left}: $k_0R=0$, which corresponds to the near-field limit of the produced axion field. \textbf{Right}: $k_0R=1$ in the $m_a\geq\omega_0$ regime. The gray dashed lines represent the boundary of the sensor medium, outside which $M_x$ is zero. The curves are normalized with the amplitude of $M_x$ at the near end $\mathbf{x}_0$.}
    \label{fig:Bx}
\end{figure*}

We calculate the amplitude of $B_x$ and $M_x$ along the $x$-axis for the setups with two cylindrical samples discussed in the main text. The results are shown in Fig.~\ref{fig:Bx}. 
For $m_a\leq\omega_0$, the setups are in the near zone of the produced axion field, which is shown in the left panel.
The right panel shows the case of $\kappa_0 R =1$ which corresponds to the maximal axion mass that can be probed with the setups.
Note that the results are independent of $R$ since $\mathcal{G}(\mathbf{x})$ only depends on dimensionless quantities $\mathbf{x}/R$, $H/R$, and $k_0R$.
The plots indicates that $B_x$ is maximal at the near end of the sensor medium, $\mathbf{x}_0$, where it is related to $M_x$ with an $\mathcal{O}(1)$ factor.

\section{Frequency and mass dependence of Sensitivities\label{sup:optimization}}

In Supplemental Material ``Likelihood Framework'', we derive the dependence of our projected sensitivity on the experimental parameters (Eq.~\eqref{upper limit scaling}).
Its dependence on $\omega_0$, $m_a$ and the geometry of the spin source is implicitly contained by $|\mathcal{G}(\mathbf{x}_0)|$ which represents the amplitude of axion-induced driving force. \
The geometrical dependence has been discussed in Supplemental Material ``Geometrical Dependence of the Signal''.
In this section, we focus on the behavior of the projected sensitivity with respect to $\omega_0$ and $m_a$.

\begin{figure*}[h!t]
        \centering
         \includegraphics[width=0.465\linewidth]{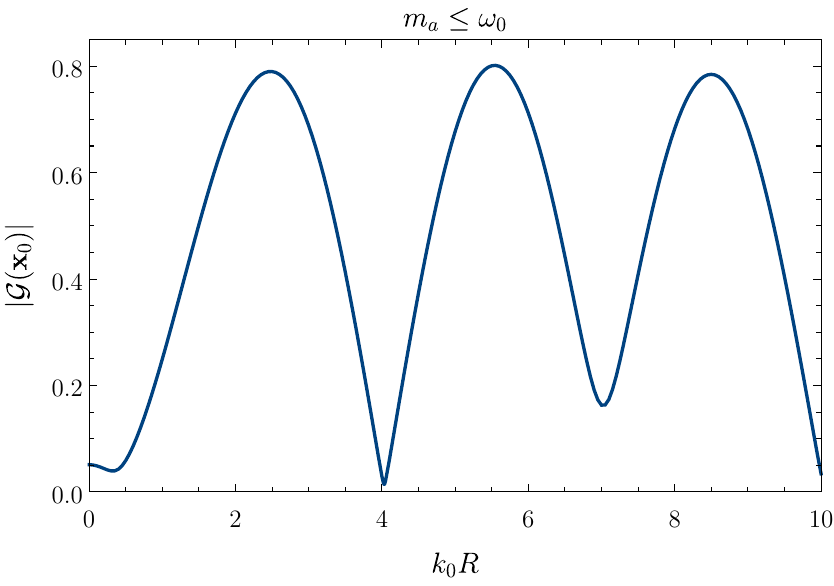}
         \qquad\includegraphics[width=0.475\linewidth]{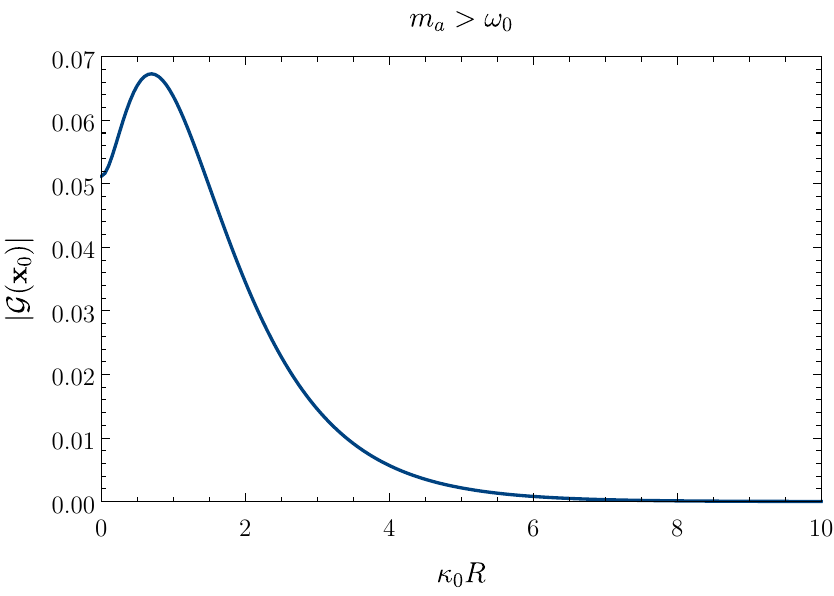}
    \caption{\textbf{Left}: $|\mathcal{G}(\mathbf{x}_0)|$ as a function of $k_0R=\sqrt{\omega_0^2-m_a^2}R$ for $m_a \leq \omega_0$. 
    \textbf{Right}: $|\mathcal{G}(\mathbf{x}_0)|$ as a function of $\kappa_0R=\sqrt{m_a^2-\omega_0^2}R$ for $m_a > \omega_0$.
    In both plots, ${\mathbf{x}_0} = (2R,0,0)$ and the height-to radius ratio of the cylindrical spin source is $H/R=3$.}
    \label{fig:G with xi}
\end{figure*}

As demonstrated by Eq.~\eqref{G(x)}, $\mathcal{G}(\mathbf{x})$ only depends on $k_0 R$ if fixing the normalized evaluation location $\mathbf{x}/R$ and height-to-radius ratio of the cylindrical spin source $H/R$.
Fig.~\ref{fig:G with xi} shows the dependence of $|\mathcal{G}(\mathbf{x}_{0})|$ on $k_0R$ (when $m_a \leq \omega_0$) and $\kappa_0R$ (when $m_a > \omega_0$), with $\mathbf{x}_0 = (2R,0,0)$ and $H/R=3$.
These plots explain the behavior of the sensitivity curves ($g_{aN}^{(\mathrm{UL})}$ as a function of $m_a$ with fixed $\omega_0$) as follows:
\begin{itemize}
    \item The value of $\omega_0 R$ sets the maximal $k_0R$ for a sensitivity curve, pinning a point $S_0$ on the curve on the left panel of Fig.~\ref{fig:G with xi}. This point represents the sensitivity for $m_a=0$. 
    \item For $0<m_a \ll \omega_0$, $k_0R$ approximately remains constant as $m_a$ increases. Thus, the corresponding point on the $\mathcal{G}(\mathbf{x}_0)$ curve stays at $S_0$ and the sensitivity remains constant. This regime corresponds to the plateau of the sensitivity curves at low masses.
    \item When $m_a \lesssim \omega_0$, $k_0R$ starts to decrease significantly and it reaches zero when $m_a = \omega_0$. As $m_a$ increases, the corresponding point $S$ moves leftward from $S_0$ along the $\mathcal{G}(\mathbf{x}_0)$ curve and the sensitivity can undergo several oscillations before $S$ arrives at $k_0R = 0$, depending on how large $(k_0R)_{\mathrm{max}}=\omega_0 R$ is. The peaks are at $k_0R \simeq2.5,\,5.5,\,8.5,$ etc. Physically, the oscillation is due to the interference of the axion field with different phases produced by different parts of the source.
    \item When $m_a = \omega_0$, point $S$ reaches $k_0R=0$ on the $\mathcal{G}(\mathbf{x}_0)$ curve on the left panel of Fig.~\ref{fig:G with xi}, which is also the $\kappa_0R=0$ point on the right panel. As $m_a$ exceeds $\omega_0$ and continues growing, $S$ moves rightward starting from $\kappa_0R=0$ on the right panel. The sensitivity at first increases, reaching a local maximum at $\kappa_0R\simeq 0.7$, and then decreases exponentially.
\end{itemize}

In the experiment setup discussed in the main text with $^{129}$Xe, $\omega_0 R \simeq 0.25 (B_0/10\mathrm{T}) (R/10\mathrm{cm})$, so our sensitivity curves in Fig~\ref{fig:sensitivity} don't go through oscillations and maintain around the near-field value $|\mathcal{G}(\mathbf{x}_0)|\simeq 0.05$ for $m_a\leq\omega_0$.
Ideally, one can tune $\omega_0R$ to the peaks ($\simeq2.5,\, 5.5,\, 8.5,$ etc.) to enhance the sensitivities in the low-mass plateaus region ($m_a\ll\omega_0$).
If future setups for the spin source could achieve  $\omega_0R\simeq 2.5$ where $|\mathcal{G}(\mathbf{x}_0)|\simeq 0.8$, the sensitivity for low masses can be strengthened by $\sqrt{0.8/0.05}=4$ times. The maximal axion mass that can be probed is set by the peak on the right panel at $\kappa_0R\simeq0.7$, larger than which, the sensitivity begins to decline exponentially.

\end{document}